\documentclass[epj]{svjour}
\usepackage{graphicx,tabularx,epstopdf,color,dcolumn,cite,amsmath}

\begin{document}
\title{Effective two-mode model in Bose-Einstein condensates versus
  Gross-Pitaevskii simulations} \titlerunning{Effective two-mode model
  in Bose-Einstein condensates vs Gross-Pitaevskii simulations}
\author{Mauro Nigro\inst{1,2}\fnmsep\thanks{\email{nigro@df.uba.ar}}
  \and Pablo
  Capuzzi\inst{1,2}\fnmsep\thanks{\email{capuzzi@df.uba.ar}} \and
  Horacio
  M. Cataldo\inst{1,2}\fnmsep\thanks{\email{cataldo@df.uba.ar}} \and
  Dora M. Jezek\inst{1,2}\fnmsep\thanks{\email{djezek@df.uba.ar}}}
\institute{Universidad de Buenos Aires, Facultad de Ciencias Exactas y
  Naturales, Departamento de F\'{\i}sica, Pabell\'on 1, Ciudad
  Universitaria, 1428 Buenos Aires, Argentina \and Instituto de
  F\'{\i}sica de Buenos Aires, CONICET-UBA, Pabell\'on 1, Ciudad
  Universitaria, 1428 Buenos Aires, Argentina } \date{\today}

\abstract{ We study the dynamics of three-dimensional Bose-Einstein
  condensates confined by double-well potentials using a two-mode
  model with an effective on-site interaction energy parameter.  The
  effective on-site interaction energy parameter is evaluated for
  different numbers of particles ranging from a low experimental value
  to larger ones approaching the Thomas-Fermi limit, yielding
  important corrections to the dynamics.  We analyze the time periods
  as functions of the initial imbalance and find a closed integral
  form that includes all interaction-driven parameters.  A simple
  analytical formula for the self-trapping period is introduced and
  shown to accurately reproduce the exact values provided by the
  two-mode model.  Systematic numerical simulations of the problem in
  3D demonstrate the excellent agreement of the two-mode model for
  experimental parameters.   }
%
%

\maketitle
\section{Introduction}

The TM model applied to double-well atomic Bose-Einstein
condensates has been extensively studied in the recent years
\cite{smerzi97,ragh99,anan06,jia08,albiez05,mele11,abad11,doublewell_a,
  doublewell_b,doublewell_c,doublewell_d,doublewell_e}.  Such a model
assumes that the condensate order parameter can be described as a
superposition of wave functions localized in each well with time
dependent coefficients \cite{smerzi97,ragh99}.  The localized wave
functions are straightforwardly obtained in terms of the stationary
symmetric and antisymmetric states, which in turn determine the
parameters involved in the TM equations of motion
\cite{smerzi97,ragh99,anan06,jia08}. The corresponding dynamics
exhibits Josephson and self-trapping regimes \cite{smerzi97,ragh99}
which have been experimentally observed by Albiez {\it et al.}
\cite{albiez05}.

The self-trapping (ST) phenomenon, which is also present in extended
optical lattices \cite{optlat_a,optlat_b,optlat_c,optlat_d,
  Anker2005,Wang2006}, is a non linear effect where the difference of
populations between neighbouring sites does not change sign during the
whole time evolution.  There is nowadays an active research on the ST
effect, which involves different types of systems, including mixtures
of atomic species \cite{stlastoplat,mele11}. Research on condensates
trapped in ring-shaped optical lattices is also a promising area given
that successful efforts has been done in their experimental
realization \cite{hen09}.  The dynamics on systems with three
\cite{trespozos2011} and four wells \cite{cuatropozos06} has been
initially investigated through multimode models that utilized {\em
  ad-hoc} values for the hopping and on-site energy parameters.
Whereas in \cite{jezek13b}, such parameters have been extracted for a
ring-shaped optical lattice with an arbitrary number of wells, by
constructing two-dimensional localized Wannier-like (WL) functions in
terms of stationary Gross-Pitaevskii (GP) states.

In recent works it has been shown that a correction in the TM model
that involves the interaction energy should be taken into account in
order to properly describe the exact dynamics
\cite{jezek13a,jezek13b}. In particular in \cite{jezek13a} an
effective two-mode (ETM) model has been developed with an interaction
parameter which has been analytically obtained in the Thomas-Fermi
(TF) limit, that completely heals this disagreement.  In the present
work, we will extend these studies for lower numbers of particles by
numerically calculating the effective parameter that enters in the
model .  Here we will analyze the double-well system with the
experimental conditions of \cite{albiez05}, where the number of
particles is $1150$, and increase such a number to show that the
correction to the on-site interaction energy parameter goes to the one
predicted in the TF regime \cite{jezek13a}.  The main goal of this
work is to assess the accuracy of the ETM model by calculating the
time periods as functions of the initial imbalance and analyze the
role of the different parameters. To this end, we shall confront the
values of the orbits periods predicted by this model to those obtained
by numerically solving the three-dimensional Gross-Pitaevskii
equations.  In particular, within the effective two-mode model
framework we derive closed expressions for the periods valid for any
imbalance value. We then develop a simple analytical approximation to
the ST period and improve the calculation of the Josephson period for
small imbalances by taking into account the parameter that involves the
density overlap between the localized states in neighbouring sites
\cite{anan06}. This correction will be of importance for the
experimental configuration of the Heidelberg group \cite{albiez05}. We
will show that the critical imbalance for the transition between the
Josephson and ST regimes predicted by our model is in good agreement
with the experimental finding in Ref. \cite{AlbiezThesis05} for the
first time.

This paper is organized as follows. In Sect.\ \ref{twomode} we
describe the double-well system and find the effective on-site
interaction energy parameter for several particle numbers. Such a
parameter is obtained from a linear approximation of the on-site
interaction energy as a function of the imbalance. We will show that
the corresponding second order term in the approximation turns out to
be much smaller and gives rise to a third order correction in the
equations of motion which can be safely disregarded. In Sect.
\ref{sec:timeperiods} we derive a closed integral form for the period
of the orbits with an arbitrary initial imbalance and obtain explicit
analytical approximations within the ST and Josephson regimes, while
the numerical results and comparisons with the GP calculations are
included in Sect. \ref{sec:NumRes}. To conclude, a summary of our work
is presented in Sect. \ref{sum} including a perspective of the
application of these methods to multiple-well systems in
configurations with high symmetries.  Finally, the definition of the
parameters employed in the equations of motion are gathered in the
Appendix.

\section{\label{twomode} Two-mode model}
We consider a Bose-Einstein condensate of Rubidium atoms confined by
the external potential $ V_{\mathrm{trap}}$ used in the experiment of
the Heidelberg group \cite{albiez05},
\begin{equation}
V_{\mathrm{trap}}({\bf r} ) = \frac{ 1 }{2 } \, m \, ( \omega_{x}^2  x^2  + \omega_{y}^2  y^2 
+ \omega_{z}^2  z^2 ) +   
V_0 \, \cos^2(\pi x/q_0)
\end{equation}
where $m$ is the atom mass, $ \omega_{x}= 2 \pi \times 66 $ Hz,
$ \omega_{y}= 2 \pi \times 78 $ Hz, and $ \omega_{z}= 2 \pi \times 90$
Hz. The lattice parameters are given by
$V_0= 2\pi\times 412 \, \hbar $ Hz and $ q_0=5.2 \,\mu$m.  The number
of particles used in the experiment is $N= 1150$, but we will also
consider particle numbers up to $N=10^5$.


\subsection{Inclusion of  effective  on-site interaction energy  effects}

In previous works \cite{jezek13a,jezek13b} we have shown that the
linear dependence on the imbalance of the interaction energy
integrated in each well gives rise to a lower effective on-site
interaction parameter. Here we will evaluate such a parameter by using
a combination of the procedures described in \cite{jezek13a,jezek13b}
and also by expanding to a higher order approximation on the
imbalance.

In doing so, we first rewrite the TM equations of motion by assuming
that the on-site interaction energy $U$ can be different in the left
($U_L$) and right ($U_R$) wells. As described in \cite{jezek13a},
$U_R$ and $U_L$ arise from introducing in the mean-field interaction
term of the GP equation a more realistic density distribution that
depends on the imbalance.  Then the GP equation projected into two
localized modes at the left and right wells yields \cite{anan06}
\begin{equation}
\hbar \frac{dZ}{dt} = - {2K} \sqrt{1-Z^2}\,\sin\varphi + I(1-Z^2)\sin 2\varphi
\label{imb1}
\end{equation}
\begin{align}
\hbar \frac{d\varphi}{dt} &= {U_R(Z)} N_R  -  U_L(Z)  N_L   + {2K}  \left[
\frac{Z}{\sqrt{1-Z^2}}\right]\cos\varphi \nonumber \\& - I Z (2+\cos 2\varphi).
\label{phase1}
\end{align}
The dynamical variables are the standard imbalance
$ Z = (N_R - N_L)/N $ and phase difference
$ \varphi= \varphi_L- \varphi_R$, where $N_R$ and $N_L$ are the number
of particles in the right and left wells, respectively. As derived in
 \cite{jezek13a} we have
\begin{equation}
U_k (\Delta N)  =  g\int d^3{\bf r}\,\,  \rho^k_N({\bf r}) \,  \rho^k_{N+\Delta N}({\bf r}),
\label{ur}
\end{equation}
where $ k=R,L$, and $\rho^k_{N}$, and $\rho^k_{N+\Delta N}$ are the
localized densities in the $k$--site for systems
with total number of particles $N$ and $N+\Delta N$, respectively.
The remaining parameters $J$, and the interaction-driven $F$ and $I$
are defined as usual \cite{smerzi97,ragh99,anan06} in terms of the
localized wave functions (see the Appendix), being $K= J+F$.

\begin{figure}
  \includegraphics[width=\columnwidth,clip=true]{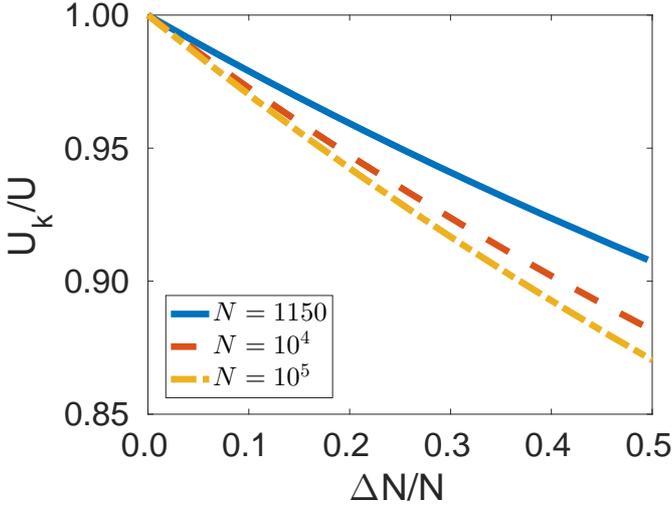}
  \caption{\label{fig:URZ}(color online) On-site interaction energy
    ratio $U_k/U$ as a function of $\Delta N/N$, for $N= 1150 $,
    $N=10^4 $, and $N=10^5 $. }
\end{figure}

Aiming at reproducing the experimental conditions of \cite{albiez05},
where the number of particles $N= 1150$ is not large enough to be in
the Thomas-Fermi regime, and thus the dependence on the imbalance of
the on-site interaction energy $U_R$ and $U_L$ cannot be analytically
calculated, we should evaluate Eq.\ (\ref{ur}). To simplify the
numerical calculation given that the wells are equal, instead of using
the localized densities in Eq.\ (\ref{ur}), we can use the alternative
method proposed in \cite{jezek13b} where only GP ground-state
densities are involved. In that work it has been shown that
$U_k(\Delta N_k)$ with $\Delta N_k=N_k-N/2$ can be evaluated as
\begin{equation}
\frac{U_k (\Delta N_k)}{U}  = \frac{ \int d^3{\bf r}\,\,  \rho_N({\bf r}) \,  \rho_{N+\Delta N}({\bf r})}
{ \int d^3{\bf r}\,\,  \rho^2_N({\bf r}) },
\label{coc}
\end{equation}
where $\rho_N$ and $\rho_{N+\Delta N}$ are the GP ground-state
densities for systems with $N$ and $N + \Delta N $ total number of
particles, respectively, being $\Delta N=2\Delta N_k$. The numerical
result of $U_k/U$ as a function of $ \Delta N /N $ has been depicted
in Fig. \ref{fig:URZ}, where it can be seen that it exhibits an almost
linear behaviour.  A second order approximation of $U_k$
\begin{equation}
\frac{U_{k}(\Delta N_k)}{U} \simeq   \  1  -   \alpha   \frac{ 2 \Delta N_k} {N}  +  \beta  \left( \frac{ 2 \Delta N_k} {N}
\right)^2
\label{Uek} 
\end{equation}
can be obtained by using a polynomial fit of the function with
parameters $\alpha $ and $\beta$.  These parameters are listed in
Table \ref{tab:1} for different numbers of particles and trapping
parameters. It is worthwhile mentioning that for the largest number of
particles considered in this work, we have taken a larger $q_0$ value
than that of the Heidelberg experiment and modified the depth of the
wells since the size of the condensate increases with the number of
particles.

\begin{table}
  \caption{\label{tab:1} Coefficients 
    $\alpha$ and $\beta$ of the quadratic fit of   $U_k/U$ as 
    a function of $\Delta N/N$ and $\gamma$  for several values of the  system 
    parameters. In  the 6$^{\mathrm{th}}$ column the factor $f_{3D}$ that    reduces the interaction energy    parameter is also given.
  }
\tabcolsep=3pt
\begin{tabular}{lccccccc}
\hline
$N$  & $q_0 (\mu\mathrm{m})$ & $V_0 (2\pi \hbar\mathrm{Hz})$ &    $\alpha$  &    $\beta$    &          $f_{3D} = 1-\alpha $   &  $\gamma$ &  
\\[2pt] \hline
$ 1150 $  & 5.2 & 412  & 0.21  &     0.06    &    0.79     & 0.064    \\[3pt]
$ 10^4  $ & 5.2  & 858  & 0.28  &     0.08  &     0.72     &   0.010      \\[3pt]
$ 10^5 $  & 8.0 & 1980  & 0.30  &     $\;\,$0.08$^{\rm a}$ &    $\;\,$0.70$^{\rm b}$   &       0.005 \\[3pt]
\hline
\end{tabular}
$^{\rm a}$ The Thomas-Fermi limit is 77/1000.\\
$^{\rm b}$ The Thomas-Fermi limit is 7/10. 
\end{table}

Introducing the expansions of $U_R$ and $U_L$ in the equation of
motion (\ref{phase1}) for the phase difference, we obtain
the on-site interaction-driven correction,
\begin{equation}
\frac{U_R(\Delta N_R)}{U} N_R  -  \frac{ U_L (\Delta N_L)}{U} N_L  = \left[(1-\alpha)  Z + \beta
 Z^3 \right] N ,
\label{resta}
\end{equation}
which yields
\begin{align}
\hbar \frac{d\varphi}{dt} &=\left[(1-\alpha)  Z + \beta 
 Z^3 \right]   U N  + {2K}  \left[
\frac{Z}{\sqrt{1-Z^2}}\right]\cos\varphi \nonumber \\ &-I Z (2 + \cos 2\varphi).
\label{phase2A}
\end{align}

We note that for all number of particles of Table \ref{tab:1} we have
$\beta Z^3 \ll (1-\alpha)Z$, and hence one can safely disregard the
term of third order in $Z$ in Eq.\ (\ref{phase2A}) in all cases. Then,
we conclude that the effective TM model can be simply obtained by
replacing the on-site interaction energy parameter $U$ by
$U_{\mathrm{eff}} = (1-\alpha) U = f_{3D} U $. For the largest number
of particles considered here, we have $f_{3D}= 7/10$ in accordance
with the analytic result obtained in the Thomas-Fermi approximation
\cite{jezek13a}, whereas for the lowest value $N=1150$, we obtain
$f_{3D}=0.79 $. Such a value does not seem to depend on the ratio of
the trap frequencies, since in \cite{abad15} the harmonic trap
frequencies are equal in the three directions and the same value of
$ f_{3D} $ was also obtained.

\subsection{ Two-mode model using the effective interaction
  parameter}

We now focus on the experimentally relevant case of $N=1150$, where we
have obtained the following TM model \cite{anan06} parameters:
$U=2.47 \times 10^{-3}\, \hbar \omega_x $,
$J=1.89 \times 10^{-2} \, \hbar \omega_x $,
$ F = 2.51 \, \times 10^{-2} \hbar \omega_x $, and
$I= 5.62\times 10^{-3} \hbar\omega_x$.  Using the results of the
previous section, we obtain
$U_{\mathrm{eff}}= f_{3D} U = 1.95 \times 10^{-3} \, \hbar \omega_x $,
with $ f_{3D}=1-\alpha =0.79 $.

In terms of the conjugate coordinates, imbalance $Z$ and phase
difference $ \varphi$, one can define the following ETM model
Hamiltonian \cite{jezek13a}:
\begin{align}
 H_{\mathrm{ETM}} (Z,\varphi) &=  \frac{1}{2} \Lambda_{\mathrm{eff}}   Z^2 - \sqrt{1-Z^2}\cos\varphi \nonumber \\
&+ \frac{\gamma}{2}(1-Z^2)(2 + \cos 2\varphi) ,
\label{eq:HETM}
\end{align}
with $ \Lambda_{\mathrm{eff}} = {U_{\mathrm{eff}} N}/{(2 K) }$ and
$\gamma=I/(2K)$.

The corresponding equations of motion are given in Hamiltonian form by
\begin{equation}
  \dot{Z} = - \frac{\partial}{\partial \varphi}H_{\mathrm{ETM}},\quad \mathrm{and}
\quad \dot{\varphi} = 
  \frac{\partial}{\partial Z}H_{\mathrm{ETM}}
\end{equation}
which yield
\begin{equation}
 \frac{dZ}{dt} = - \sqrt{1-Z^2}\,\sin\varphi + \gamma(1-Z^2)\sin 2\varphi
\label{imb}
\end{equation}
\begin{equation}
 \frac{d\varphi}{dt} =  \Lambda_{\rm \mathrm{eff}} Z + \left[
\frac{Z}{\sqrt{1-Z^2}}\right]\cos\varphi -\gamma Z (2+\cos 2\varphi),
\label{phase2B}
\end{equation}
where the time $t$ is given in units of $\hbar/2 K $.

The separatrix between Josephson and ST orbits on the phase portrait
$( Z, \varphi )$ has a critical imbalance $Z_c$ determined by the
condition $H(Z_c,0)=H(0,\pi)$, which yields
\begin{equation}
 Z^{\mathrm{ETM}}_c = 
2\frac{\sqrt{\Lambda_{\mathrm{eff}}-3\gamma - 1 }}{\Lambda_{\mathrm{eff}} -3\gamma}     \, .
\label{eq:ZcETM}
\end{equation}
Using $ \Lambda_{\mathrm{eff}} = 25.5$ we obtain a critical imbalance
$ Z_c^{\mathrm{ETM}} = 0.389$ which is much closer to that numerically
found, $ Z_c^{\mathrm{GP}} = 0.39$, than the value
$ Z_c^{\mathrm{TM}} = 0.347 $ obtained with the bare
$ \Lambda = 32.27 $ from the TM-model version improved by Ananikian
\textit{et al.} \cite{anan06}. We also note that the effect of
$\gamma$ is negligible in the $Z_c$ calculation. The numerical value of
$Z_c^{\mathrm{GP}}$ was obtained by analyzing the time evolutions of
the GP equation with different initial conditions as done in 
\cite{mele11}. In contrast to previous approximations, the value of
$Z_c^{\mathrm{ETM}}$ compares very well with the experimental finding
of the Heidelberg group as indicated  in \cite{AlbiezThesis05}.

We can estimate the relative deviation between the ETM
and TM models as
\begin{equation}
 \frac{\Delta Z_c}{Z_c^{\mathrm{TM}}} \simeq \frac{1}{ \sqrt{f_{3D}}} - 1\,, 
\label{error}
\end{equation}
which goes from $0.13$ for $N=1150$ to $0.2$ for the largest $N$
considered.

\section{\label{sec:timeperiods} Two-mode model  periods}
\subsection{Exact determination}
The time periods of orbits in both the TM and ETM models can be
obtained for any initial imbalance $Z_i$ and phase difference
$\varphi_i$. For a classical Hamiltonian system such as that described
by  Eq.\ (\ref{eq:HETM}) we can obtain the period $\tau$ from the line
integral over a given trajectory,
$\tau=-\oint 1/(\partial H/\partial \varphi) dZ$ \cite{Huang2013}.
Following this approach, an expression which does not include the
parameter $\gamma$ was previously obtained in \cite{Fu2006,Huang2013}
for the TM model. Here we extend that result and show that an
expression incorporating $\gamma$ can also be achieved, demonstrating
that this correction may be important in the Josephson regime.

The period $\tau$ of a given trajectory can be calculated from the
integral $\tau=\oint (1/{\dot{Z}})dZ$ where
$\dot{Z}$ is given by \ (\ref{imb}).
The relation between $Z$ and $\varphi$ is obtained for a given energy
$E$ by setting $H(Z,\varphi)=E$, yielding a quadratic equation for
$\cos{\varphi}$ with the solution
$\cos\varphi=\frac{1}{2{\gamma}\sqrt{1-Z^2}}(1-\sqrt{Y})$, where
$Y=1-2\gamma[(\Lambda-\gamma)Z^2-2E+\gamma]$. Taking this into account
the time period is given by
\begin{equation}
  {\tau}(Z_i,{\varphi}_i)=2\int_{Z_m}^{Z_M}dZ\frac{1}{\sqrt{Y}}\frac{1}{\sqrt{1-Z^2-\frac{1}{4{\gamma}^2}(1-\sqrt{Y})^2}}
\label{eq:tau}
\end{equation}
where $Z_m$ ($Z_M$) is the minimum (maximum) imbalance reached by the
system. The values of $Z_m$ and $Z_M$ are obtained from the phase
diagram that emerges by setting $H=E$, and have different expressions
depending on the regime. In the Josephson regime ($Z_M < Z_c$) the
conditions are $H(Z_i,\varphi_i)=H(Z_M,0)=H(Z_m,0)$ with
$Z_M>0, Z_m=-Z_M$, which give
\begin{equation}
  Z_{m\atop M}=\mp\sqrt{\frac{2}{A^2}\left[AB-1+\sqrt{C}\right]},
\end{equation} 
where $A=\Lambda-3\gamma$, $B=E-{3\gamma}/{2}$ and
$C=(AB-1)^2-A^2(B^2-1)$.  On the other hand, in the ST regime, taking
into account that the phase diagram is symmetric under the inversion
of $Z_i$ we restrict the domain of $Z_i$ to $Z_i>0$. In this case the
conditions read $H(Z_i,\varphi_i)=H(Z_M,0)=H(Z_m,\pi)$ valid for
$Z_M > Z_c$, which yield
\begin{equation}
Z_{m\atop M}=\sqrt{\frac{2}{A^2}\left[AB-1\mp\sqrt{C}\right]}.
\end{equation}
This formulation can also be used for the ETM model by replacing
$\Lambda$ by $\Lambda_{\mathrm{eff}}$. It is worthwhile mentioning
that the expression (\ref{eq:tau}) for $\gamma=0$ can be
written in terms of the complete Elliptic integral of the first kind
$\mathcal{K}(k)$, as shown previously in  \cite{ragh99} by directly
integrating the equations of motion for $Z(t)$ and $\varphi(t)$.

\subsection{Approximate expressions}
Even though the above formalism provides a closed integral form for
the time periods amenable to a numerical calculation, both in the
Josephson and ST regimes, it is also useful to derive analytical
expressions in specific limits.  In the case of small oscillations, by
retaining only quadratic terms in the Hamiltonian, Eq. (\ref{eq:tau})
can be straightforwardly integrated and we recover the expressions
given by the standard formula in \cite{smerzi97,mele11} with the
inclusion of $\gamma$ \cite{anan06}. Replacing $U$ by
$U_{\mathrm{eff}}$, one thus obtains the ETM model period,
\begin{equation}
T^{\mathrm{ETM}}_{so}= \frac{\pi \hbar }{ K \sqrt{(\Lambda_{\mathrm{eff}} + 1-3\gamma)(1-2\gamma)}}  \,,
\label{eq:tpeqosc}
\end{equation}
which yields $T^{\mathrm{ETM}}_{so}= 14.91 \, \omega_x^{-1} $ in
contrast to $ T_{so}^{\mathrm{TM}}= 13.29 \, \omega_x^{-1} $ obtained
using the bare $\Lambda$ value. We remark that an important correction
is also provided by the parameter $\gamma$. This correction diminishes
for increasing $Z_i$, and it does not affect sizeably either the
critical imbalance $Z_c$, or the time periods in the ST
regime.

In the ST regime one can also derive a limiting approximation for the
time period valid for large $\Lambda$. In this case, we can neglect
$\gamma$ and take the first-order approximation of the function
$\mathcal{K}(k)$ around $k=0$. This yields the analytical expression
for the time period $\tau_0$ 
\begin{equation}
\tau_0 = \frac{\hbar}{2K}\frac{2\pi}{\Lambda Z_i}.
\label{eq:tau0}
\end{equation}
A higher-order approximation of $\mathcal{K}(k)$ could also be
employed to increase the accuracy of $\tau_0$, but since one should
retain an important number of terms to achieve a noticeable
improvement, this procedure become quite cumbersome thus relegating
the simplicity of Eq.\ (\ref{eq:tau0}).

However, a simple analytical formula that improves $\tau_0$ can be
developed in the ST regime by performing some approximations directly
in the equations of motion.  We will keep assuming a large interaction
parameter $ \Lambda $ and neglect the parameter $\gamma$, as it does
not contribute to any significant change in the predictions of this
regime. Considering the imbalance performs oscillations around a
positive mean value and using $ \Lambda \gg 1$,  (\ref{phase2B})
can be approximated by,
\begin{equation}
 \frac{d\varphi}{dt} =  \Lambda Z + \left[
\frac{Z}{\sqrt{1-Z^2}}\right]\cos\varphi \simeq \Lambda  Z \simeq \Lambda  Z_0,
\label{Iphaseap}
\end{equation}
where $ Z_0 = \overline{Z(t)} $ denotes the mean value of the time
dependent imbalance, and we have used that the second term of
Eq.\ (\ref{Iphaseap}) averages approximately to zero. Then, assuming
$\varphi(0)=0$ we  obtain,

\begin{equation}
 \varphi(t) =   \Lambda  Z_0 t, 
\label{Iphaseapt}
\end{equation}
which replaced in  Eq.\ (\ref{imb}) with the further assumption that
$\sqrt{1-Z^2}\simeq \sqrt{1-Z_0^2}$ yields,

\begin{equation}
 \frac{dZ}{dt} = - \sqrt{1-Z_0^2}\,\sin( \Lambda  Z_0 t). 
\label{dimbst}
\end{equation}

Integrating the last expression with respect to time and considering
the initial value $Z(0)= Z_i$, we finally obtain for small $Z_0^2$ 

\begin{equation}
  Z(t) = \left(1-\frac{Z_0^2}{2}\right)\frac{\cos( \Lambda  Z_0 t)}{\Lambda  Z_0}- \left(1-\frac{Z_0^2}{2}\right)\frac{1}{\Lambda  Z_0} + Z_i . 
\label{imbst}
\end{equation}

Furthermore, to be consistent with $Z_0$ being the mean value of
$Z(t)$, we impose

\begin{equation}
 Z_0 = - \left(1-\frac{Z_0^2}{2}\right)\frac{1}{\Lambda  Z_0} + Z_i ,
\label{conz0}
\end{equation}
which yields a quadratic equation for $Z_0$ with the following
solution for $\Lambda\gg 1$,

\begin{equation}
 Z_0 = \frac{Z_i }{ 2 }  \left[ 1 \pm \sqrt{1- \frac{4}{\Lambda Z_i^2}} \, \right]\, .
\label{z0}
\end{equation}

Given that we are assuming a ST regime, which implies that
$Z(t)$ from  Eq.\ (\ref{imbst}) should not change sign during the
evolution, we discard the minus sign in front of the square root in
Eq.\ (\ref{z0}).
%
%
Therefore, using  Eq.\ (\ref{z0}) we can estimate the ST period
$T_{st}=2\pi/(\Lambda Z_0$) as,
\begin{equation}
T_{st}=  \frac{ Z_i  \pi \hbar }{ 2 K }\left( 1 - \sqrt{1- \frac{4}{\Lambda Z_i^2}}\right) ,
\label{eq:tstbuenaa}
\end{equation}
which will be expressed in units of $\omega_x^{-1}$.  The above
equation can also be used by replacing $\Lambda$ by
$ \Lambda_{\mathrm{eff}}$ to better take into account the effective
interaction effects. For example, for an initial imbalance $Z_i=0.45$
it yields $ T^{\mathrm{ETM}}_{st}= 8.42 \, \omega_x^{-1} $ and
$ T_{st}^{\mathrm{TM}}= 6.05 \, \omega_x^{-1} $ in comparison with
that obtained with the GP simulation,
$T_{st}^{\mathrm{GP}}=8.54\,\omega_x^{-1}$.

\section{\label{sec:NumRes}Numerical results}
\begin{figure}
\includegraphics[width=\columnwidth,clip=true]{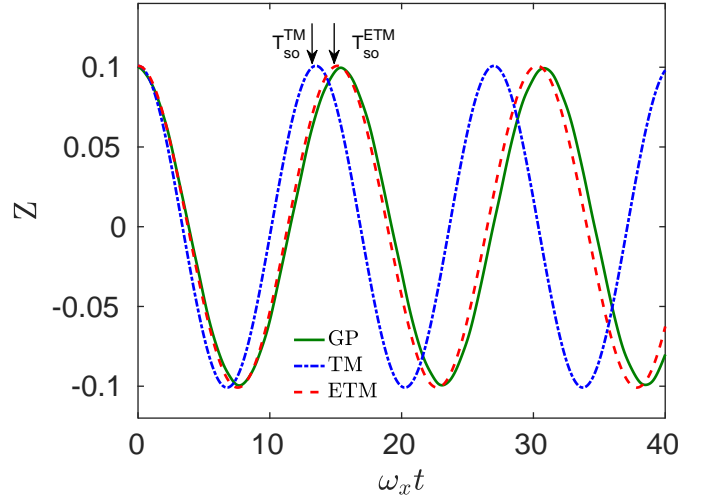}
\caption{\label{fig:gpeTM_01}(color online) Time evolution of an
  initial imbalance in the Josephson regime using the GP equation, the
  TM and ETM models for the initial condition $Z_i=0.1$ and
  $\varphi_i=0$. The vertical arrows indicate the small-oscillation
  period estimates for both models, Eq. (\ref{eq:tpeqosc}). }
\end{figure}

\begin{figure}
\includegraphics[width=\columnwidth,clip=true]{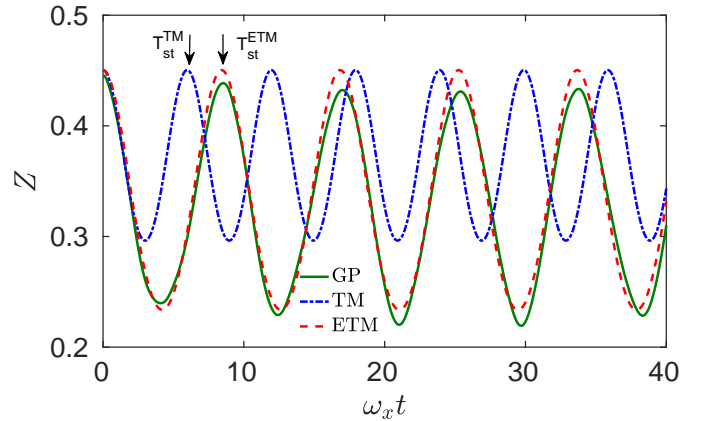}
\caption{\label{fig:gpeTM_045}(color online) Time evolution of an
  initial imbalance in the ST regime using the GP equation,
  the TM and ETM models for the initial condition $Z_i=0.45$ and
  $\varphi_i=0$. The vertical arrows indicate the ST period
  estimates arising from Eq.\ (\ref{eq:tstbuenaa}) for both models. }
\end{figure}

Aiming at testing the validity of the model equations, we have
numerically solved the GP equation using a second order in time,
split-step spatial Fourier operator \cite{NR,bao2003} with up to
$512\times 512\times 256$ grid points and time steps down to
$\Delta t=5\times10^{-5}\omega_x^{-1}$. In Figs. \ref{fig:gpeTM_01}
and \ref{fig:gpeTM_045} we show the GP time evolutions for initial
imbalances in the Josephson and ST regimes, respectively, as compared
to those given by TM models using the bare $ \Lambda $ and the
effective $ \Lambda_{\mathrm{eff}}$ values.  It becomes clear that the
effective approach reproduces the GP results much better than the bare
TM model in both regimes.  We also notice that the small-oscillation
period (\ref{eq:tpeqosc}) calculated from the effective interaction
parameter is a much better estimate and the same holds for the period
estimates given by Eq.\ (\ref{eq:tstbuenaa}) in the ST regime.

\begin{figure}[ht]
\includegraphics[width=\columnwidth,clip=true]{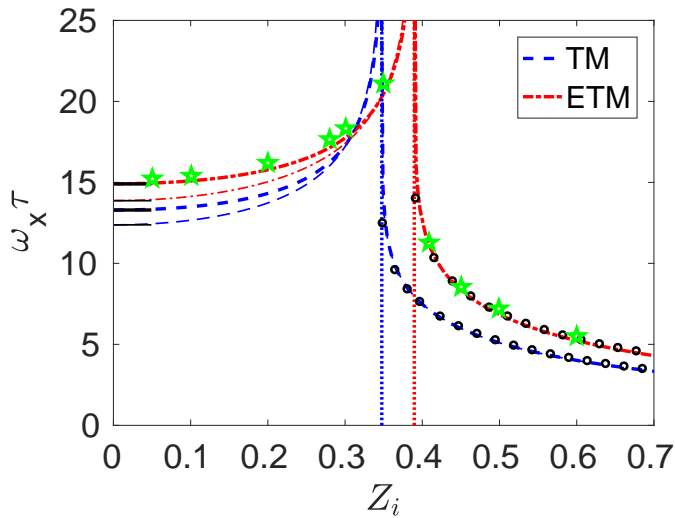}
\caption{\label{fig:periodos} (color online) Trajectory periods as
  functions of the initial imbalance $Z_i$ for the TM (dashed blue
  lines) and ETM (dash-dotted red lines) models according to Eq.\
  (\ref{eq:tau}) with $\varphi_i=0$. Thinner lines correspond to
  calculations neglecting $\gamma$. The vertical lines mark the
  critical imbalance $Z_c$, while the circles correspond to Eq.\
  (\ref{eq:tstbuenaa}) for the TM and ETM models and the horizontal
  solid lines correspond to the small-oscillation approximations. The
  stars indicate the periods obtained from the full 3D GP simulation.}
\end{figure}

In Fig. \ref{fig:periodos} we compare the time periods as a function
of the imbalance using the TM and ETM models together with several
periods obtained from GP simulations.  We also plot with empty circles
$T_{st}(Z_i)$ from Eq.\ (\ref{eq:tstbuenaa}) and with horizontal lines
$T_{so}$ given by Eq.\ (\ref{eq:tpeqosc}), both for the TM and ETM
models.  We notice that the predictions for both the ST and the
small-oscillation periods are highly accurate within both two-mode
models, and that the ETM results agree well with the full GP
calculation. We have also included in Fig. \ref{fig:periodos}
calculations neglecting $\gamma$ (depicted in thinner lines), so as to
emphasize that for the experimentally relevant case of $N=1150$ the
inclusion of the parameter $\gamma$ also yields a sizable correction
to the Josephson periods.  On the other hand, for smaller overlaps
between the densities of the localized states, the factor $\gamma$ is
substantially reduced (cf. Table \ref{tab:1}) and thus it does not
play any significant role in determining these periods.

\begin{figure}[ht]
\includegraphics[width=\linewidth,clip=true]{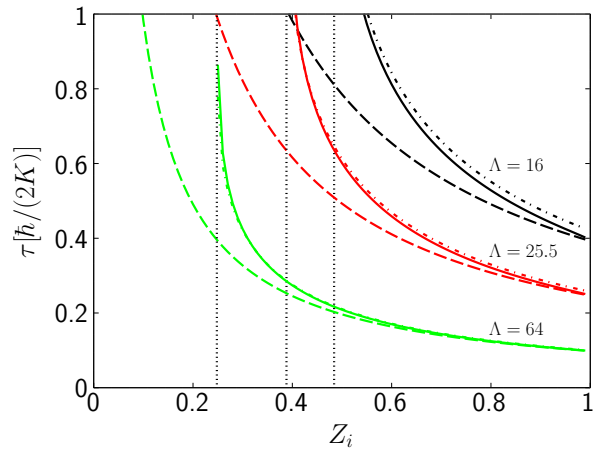}
\caption{\label{fig:periCompa}(color online) Comparison of the time
  periods $\tau$ (in units of $\hbar/(2K)$) in different
  approximations for $\Lambda=16, 25.5$, and $64$. The solid, dashed,
  and dash-dotted lines correspond to the exact results
  (\ref{eq:tau}), the approximation $\tau_0$ (\ref{eq:tau0}), and
  $T_{st}$ given by Eq.\ (\ref{eq:tstbuenaa}), respectively. The vertical
  dotted lines mark the critical imbalance $Z_c$ for each value of
  $\Lambda$.}
\end{figure}

We also compare in Fig. \ref{fig:periCompa} the exact results for
$\gamma=0$ in the ST regime with the value of $\tau_0$ given by 
Eq.\ (\ref{eq:tau0}), and with $T_{st}$, Eq.\ (\ref{eq:tstbuenaa}). In
particular, we show the results for $\Lambda=16, 25.5$, and $64$,
where it may be seen that our estimate, $T_{st}$, provides a simple
and improved overall approximation around an extended region in
$Z_i$. For lower values of $\Lambda$ the assumption $\Lambda \gg 1$
breaks down and hence both approximations becomes less accurate.  For
larger values both estimates get closer to the exact result, while our
prediction is able to quantitatively describe the exact curve closer
to $Z_c$ much better than $\tau_0$. For values above
$\Lambda\simeq 10^3$ despite the error is substantially reduced in
both approximations, $T_{st}$ still improves the period calculation
over $\tau_0$.

\section{\label{sum} Summary and concluding remarks}

We have studied the dynamics of three-dimensional Bose-Einstein
condensates using a two-mode model with an effective on-site
interaction parameter and compare it to the full 3D Gross-Pitaevskii
simulations.  We demonstrate that the periods of the orbits for
two-mode models with arbitrary initial conditions can be written as a
closed integral form which takes into account the effect of the
overlap between the localized densities through the parameter
$\gamma$. We show that this interaction-driven effect is specially
important in the Josephson regime for the experimental conditions of
\cite{albiez05}. Furthermore, based on the dynamical equations for the
populations and phase differences in each well, we have derived a
simple analytical formula for the period in the self-trapping regime,
which accurately reproduces the exact integral expression of the
two-mode model and correctly describes Gross-Pitaevskii simulation
results for large on-site interaction energy parameters.

The three-dimensional numerical simulations prove that the precise
determination of the effective on-site interaction energy parameter is
essential to correctly reproduce the GP results and thus to calculate
accurate estimates of the time periods.

The present study opens the possibility to the application of the
effective two-mode model and the time period expressions to
multiple-well systems with symmetric initial populations. In such
cases, the dynamics can be characterized by a single imbalance and a
phase difference in terms of which the two-mode Hamiltonian can be
easily furnished.  Studies in such direction are currently underway
for a four-well system.

\begin{acknowledgement}
 This work was supported by CONICET and Universidad de Buenos
Aires through grants PIP 11220150100442CO and UBACyT 20020150100157,
respectively.
\end{acknowledgement}
\subsection*{Author contribution statement}
All authors contributed equally to the paper.

\appendix
\setcounter{equation}{0}
\def\theequation{A.\arabic{equation}}
\def\thesection{\appendixname\ \Alph{section}.}
\section{Multimode Parameters \label{sec:parameters}} 

The parameters of the TM model are defined as

\begin{equation}
J= -\int d^3{\bf r}\,\, \psi_R ({\bf r}) \left[
-\frac{ \hbar^2 }{2 m}{\bf \nabla}^2  +
V_{\rm{trap}}({\bf r})\right]  \psi_L({\bf r})
\label{jota0}
\end{equation}
\begin{equation}
U= g \int d^3{\bf r}\,\,  \psi_R^4({\bf r})
\label{U0}
\end{equation}
\begin{equation}
F= -  g N \int d^3{\bf r}\,\,  \psi_R^3({\bf r})
 \psi_L ({\bf r})
\label{jotap0}
\end{equation}
\begin{equation}
I= g N \int d^3{\bf r}\,\,   \psi_R^2({\bf r}) \,  \psi_L^2({\bf r}) . 
\label{ijotap0}
\end{equation}
where $\psi_R(\mathbf{r})$ and $\psi_L(\mathbf{r})$ are the localized
modes at the right and left sides, respectively. As usual the left
(right) mode is obtained from the sum (difference) of the
lowest energy symmetric and antisymmetric stationary order parameters
obtained from the GP equation. The interaction-driven parameters $F$
and $I$ were first defined in \cite{anan06} and later analyzed in
\cite{jia08}.

Together with the calculation of these parameters through the
preceding definitions, we have applied also the alternative method
outlined in  \cite{jezek13b}, finding an agreement between both
procedures with a precision higher than 99\%. In particular, we note
that the difference of energies between antisymmetric and ground
states of the TM model defines the hopping parameter $K$
\cite{leblanc11}.

\end{document}